\documentclass{elsart}
\usepackage{epsfig}

\begin{document}

\begin{frontmatter}

\title{Virtual volatility}
\author{A. Christian Silva\corauthref{presentAddress}}
\ead{silvaac@evafunds.com}
\author{, Richard E. Prange}
\ead{prange@glue.umd.edu }

\corauth[presentAddress]{Present address: EvA Funds, 456 Montgomery
St, 8th floor, San Francisco, CA 94104}


\address{Department of Physics, University of Maryland, College Park,
  MD 20742-4111, USA}

\begin{abstract}
We introduce the concept of virtual volatility. This simple but new measure
shows how to quantify the uncertainty in the forecast of the drift component
of a random walk. The virtual volatility also is a useful tool in
understanding the stochastic process for a given portfolio. In particular, and
as an example, we were able to identify mean reversion effect in our
portfolio. Finally, we briefly discuss the potential practical effect of the
virtual volatility on an investor asset allocation strategy.
\end{abstract}


\end{frontmatter}

\section{Introduction}

\emph{Return on investment} is a central concept for investors. \ We here
consider the simplest case of buying and holding a stock which, to make things
easy, has no dividends or stock splits during the holding period. We assume a
buy and hold strategy which ends at a horizon time $T$ later. In this paper,
we are primarily interested in horizon times $T\,\ $of the order of a year.
[We do not treat the important issue of diversification and correlations
between the returns of different stocks. To some extent, this can be treated
by considering the entire portfolio in place of a single stock.]

The return can be defined as
\begin{equation}
R_{T,t}=\ln(S_{T+t}/S_{t}) \label{RT}%
\end{equation}
where $S_{t}$ is the cost of the stock per share at time $t$ in the buy
transaction, and $S_{T+t}$ is the amount to be realized per share at the time
$T+t$ of the selling transaction. This is easy enough to calculate after the
transactions are history. That result is called the historical return.

Some idea of what the return is \emph{expected to be} when the decision is
made to buy or not buy the stock is essential however. \ The problem is to
\emph{predict}, \emph{ex ante, }the stock price at a time in the future. It is
universally accepted that there is a stochastic component to the changes in
price. This leads inexorably to the idea that the best that can be hoped for
is a prediction of the \emph{probability distribution }\ [PDF] of returns at
time $T+t$, which we call $p_{t}(R|T)$. \ It is fairly generally accepted that
for most stocks the distribution is likely to be more or less bell-shaped,
with a `center' parameter and a `width' parameter. \ `Skewness' and `fatness
of tail' parameters are also interesting but shall not concern us here.

We arrive at the conclusion, as many others have done, that for horizon times
of a year, this distribution of returns can be taken to be \emph{normal,
}[log-normal for the stock], at least as a reasonable starting point. This
means that the main two parameters are the width and the center parameter. The
latter is generally taken to be the \emph{expected }or \emph{mean return,
}which we define as
\begin{equation}
\mu_{T}=E_{t}(R_{T})=\int dR~R\,p_{t}(R|T)\label{mu}%
\end{equation}
The width $\sigma_{T}$ is usually determined from $v_{T}=\sigma_{T}^{2}$ , the
variance of the distribution
\begin{equation}
v_{T}=E_{t}\left(  \left[  R_{T}-\mu_{T}\right]  ^{2}\right)  \label{var}%
\end{equation}
[An equivalent and perhaps more popular definition of the `mean return' is
$\tilde{\mu}_{T}$, defined as $\tilde{\mu}_{T}=\ln\left(  E\left(
S_{T+t}/S_{t}\right)  \right)  .$ If the PDF is lognormal, $\tilde{\mu}%
_{T}=\mu_{T}+\frac{1}{2}v_{T}.]$ \

The dependence of the PDF on $t$ emphasizes that the distribution is that
\emph{predicted} at time $t.$ This PDF is exceptionally important because it
is the one that is \emph{acted on} by the investor. It is conditional on past
history, including the price $S_{t}$ as well as all other information and
theory that the investor is able to bring to bear. Thus, in a sense there is a
different PDF for every investor. Of course, it's probably true that only a
small fraction of investors think in terms of PDF's. What follows is in the
nature of advice to those few investors on how to improve systematically their PDF's.

The expected return is also called the drift. The latter terminology comes
from the random walk model discussed further below. The parameter $\sigma_{T}$
is one among several distinct definitions of the \emph{volatility. \ }\ The
volatility is almost always, in spite of cogent criticisms, taken to be
practically synonymous with the concept of risk. It is obviously a sort of
inverse predictability, the larger the volatility, the less the horizon price
is predictable. The whole point of this paper is to discuss this volatility
and its relationship to the concept of drift.

This class of distributions of stock prices in the future must be
sharply distinguished between somewhat similar distributions in
common use. One example is the \emph{unconditional distribution} of
Lo and Wang, \cite{LoWang},LW, which `fixes' the parameters,
(including $\mu$ and $\sigma)$ at their `true' values. [Quotes are
as in LW]. Another is the `risk neutral world' PDF used in option
pricing theory, which is usually stated to be the distribution of
the price of the underlying stock  at the future time $t+T,$ with
the drift parameter $\tilde{\mu}_{T}$ replaced by the risk-free
return $rT.$ We think, that there is an explicit formula for what to
use as the option pricing volatility parameter giving a result
$\sigma_{BS}$ related to the Black-Sholes theory
\cite{BS,Bodie,Bjork}. Closely related to this is the risk free
world distribution in which the volatility parameter is the
`implied' volatility, $\sigma_{I}$ extracted from empirical option
prices \cite{Bjork}. The implied volatility in effect incorporates a
number of corrections to that of Black-Sholes. Although these
volatilities are correct for option pricing purposes, we show that
\emph{neither} of these volatilities is correctly used as the
standard deviation of $p_{t}(R|T).$

\begin{figure}
\centerline{\epsfig{file=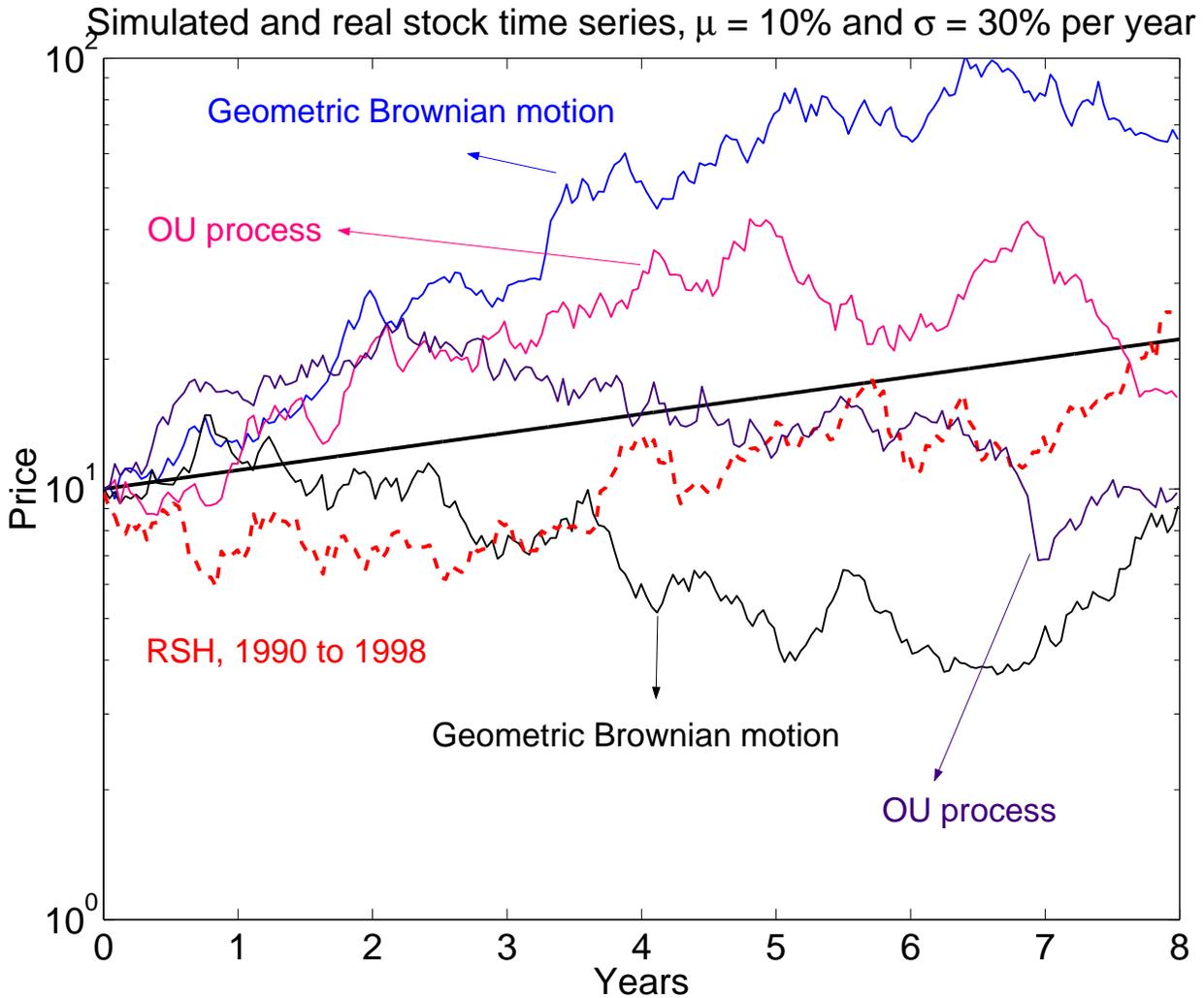,width=1\linewidth,angle=-90}}
\caption{\footnotesize\sf Stock price time series for simulated and
  one real stock (RSH, Radio Shack) with similar mean return and
  volatility of $\sigma = 0.3$ and $\mu = 0.1$ per year. The smallest
  time step for the simulation is $1$ day. The solid black line
  represent the mean drift of $0.1$ per year. Note that it is not
  possible to extract the correct mean return by simply looking at one
t  simulated time series. Therefore it is questionable to assume
that RSH
  comes from a random process presenting $10\%$ mean return. All we can say is that RSH had a realized return of $10\%$ for the period in question.
It is also
  difficult to distinguish between different stock process. For
  comparison we present $2$ simulated processes. One is the traditional
  geometrical Brownian motion, the second is an OU process with a
  target price of $15$ dollars after $1$ year. Both are
  similar if we look at few time series.}
\label{fig:simul}
\end{figure}

Suppose we have the daily price history of a single stock over a year.
Assume for the sake of argument that this history is generated by a Monte
Carlo program with constant parameters, whose numerical values are in the
range expected for stocks, but we don't know the parameters (Fig. \ref{fig:simul}). The exercise is
to estimate the parameters from the data. We will argue that we can determine
the width parameter $\sigma_{BS}$ fairly well from the data of a single year.

From a single year's data, the best we can do to  estimate the expected return
is  $\mu_{T}\approx R_{T}\pm\sigma_{BS}$, which will get the sign wrong fairly
often (Fig. \ref{fig:simul}). It is certainly known that it is impossible to measure the historical
mean drift very well. Decades of data would be needed, even on the dubious
assumption that the drift and other parameters are constant over those
decades\cite{Black1995}. Predicting the expected
return parameter is even harder. It is nevertheless routine to act as if the
drift parameter is `fixed' at its `true' value.

To obtain the parameters of the yearly distribution, the Monte Carlo for a
year could be run many times. This is something that \emph{cannot} be done for
an actual security. If the program were for a standard random walk, as is
assumed for simplicity by Black-Sholes and the efficient market hypothesis,
that would determine $\mu_{T}$ and it would be found that the yearly width is
$\sigma_{BS.}$ If the program were for a mean reverting Ornstein-Uhlenbeck [OU]
process the result would be a width $\sigma_{OU}<\sigma_{BS},$ and the result
would depend on another parameter $\gamma,$ the rate of reversion to the mean.
We estimate crudely below that $\gamma^{-1}$ is of the order of 1 year for
actual stocks. Other stochastic models are by no means ruled out
empirically.

The main, albeit quite intuitive, result of this paper is as
follows: Clearly, since one doesn't know the drift very well, one
should make some sort of average over the reasonable predictions of
that parameter. This leads to a systematic result: the volatility of
a predicted stock price, as defined by Eq.(\ref{var}) is
\emph{greater} than the volatility of the unconditional
distribution, and also greater than the volatilities $\sigma_{BS}$
or $\sigma_{I}.$ We call this enhanced volatility $\sigma_{T},$ as
above, and we estimate empirically, using $304$ companies which are
part of the S\&P500 \footnote{We find the enhanced volatility from a
point of view of a momentum investor. Furthermore, we adopt a
cross-sectional approach, which assumes that all $304$ stocks differ
only by their standard deviation and mean. Ranges for the virtual
volatility change slightly with different data sets and periods.
Using $408$ names of the S\&P 500 from Oct.1993 to Oct.2002, we have
1.3 to 1.42. Please refer to section \ref{sec:data} for more
details.},
 that $\sigma_{T}%
\approx\left(  1.26-1.34\right)  \sigma_{BS}.$ We call this 26-34\%
enhancement, the \emph{virtual }volatility, which comes, not because of a
stochastic process, but because of our inability to predict or even measure
accurately, the mean return. In other words, this is \emph{not} the standard
deviation of any given process, but rather that of an ensemble of processes,
an ensemble which is virtual because it exists only in our heads.

Finally we point out a remarkable and often overlooked consequence to investors of the existence of the virtual volatility.


\section{Theoretical and Empirical Distributions of Returns}


In the stock market, where every day billions of dollars generate
gigabytes of data, one might think that reasonable empirical
estimates of the distribution of such a fundamental quantity as the
return would be commonplace. This is true, but only for short
horizon times. For example, suppose $T$ = 1 hour. It is then
plausibly assumed that, except for stochastic processes, the return
during each hour is the same. So, one could let $t$ in Eq.(\ref{RT})
run over each hour for a couple of years to obtain a large number of
exemplars. The hourly \emph{mean return }is in one sense known very
well. Namely, it is much smaller than the hourly volatility.
However, it not known well enough to use it to estimate very well
the annual mean return, even on the assumption that the hourly
return is constant.

Of course, one should not really include every hour, just every trading hour.
Or perhaps the returns during the final hour of each trading day, or maybe
the last hour on Fridays or before holidays could be examined. These are
conditional distributions, all hourly, which are known to be significantly
different from one another. It is for some purposes quite acceptable to
average over all the constraints (except for not including nontrading hours).

It is known that for short horizons, up to a week, say, that the
\emph{shape} of the empirical returns depends significantly on $T.$
At minutes, there are power law tails and significant correlation
between successive minutes \cite{Lo,Stanley,SPY}. At hours or days,
the main part of the return is approximately exponential, with
little apparent correlation between one day and the next. At still
longer times the return distribution evolves in the direction of
normality \cite{Lo,Stanley,SPY}. However, for these longer times,
quarters, years, decades, one should not in general assume that all
the underlying parameters are constant for a given stock for a
sufficient length of time that an empirical distribution obtained
from the time series of a single stock is well defined. Certainly,
for horizons much longer than a year, the supposition that the
parameters are constant is hard to justify.

The reason for this trend to normality is of course the Central Limit Theorem.
There are many small reasons for the changes in price of a stock. These
reasons have fairly short time scale parameters. In the course of a year, the
fluctuations with short time characteristics get averaged out.

To make progress, one needs to invoke some theory. We assume that the
stock follows a generalized random walk with \emph{finite} time increment
$\Delta t$. It is usual in academic economic theory to take $\Delta t$ as an
infinitesimal, in order to connect to the beautiful theory of stochastic
differential equations, but, as we have just seen, there is much going on at
short times that only indirectly affects the yearly behavior. So we think of
$\Delta t$ as being some interval like a day, short compared with $T,$ but
long enough to minimize intraday effects.

Let $s_{t}=\approx\left(  S_{t+\Delta t}-S_{t}\right)
/S_{t}=\Delta S_{t}/S_{t}.$ Then we assume something like Brownian motion, or
a random walk,
\begin{equation}
s_{t}=\mu_{t}\Delta t+\sigma_{t}\Delta t^{1/2}W_{t}. \label{bm}%
\end{equation}
Here $W_{t}$ is a random number of mean zero and unit variance. The daily
drift $\mu_{t}\Delta t$ is small, and for our purposes nonstochastic, but it
can be a function of time and past history. The daily standard deviation
$\sigma_{t}\Delta t^{1/2}$ can be constant, time dependent, or even mean
reverting stochastic with correlations \cite{SPY}, but we assume that maximum correlation times for the daily volatility are such that most of the effects of volatility fluctuations are averaged out
over times of order a year. We assume that $\left(  \mu_{t}\Delta t\right)
^{2}<<\sigma_{t}^{2}\Delta t.$ \ [This is based on the estimate that the drift
for a year is typically in the range 10-20\%, while the volatility for a year
is 20-40\%. \ Let $\Delta t$ be about $1/252$ years. Then $\left[
.15/252\right]  ^{2}=4\times10^{-7}<<4\times10^{-4}=.3^{2}/252.$] This is a very
fundamental inequality. It says that at short times, $\Delta t\approx$ a
day, the drift is so small that for most purposes it can be neglected.

The usual Black-Sholes theory of option pricing takes $\mu_{t}$ and
$\sigma_{t}$ as constants. This is unnecessary and we can consider
them to be time dependent \cite{LoWang}. We adopt units of time such
that $\Delta t=1$ and then $T$ is large, e. g. one year $\approx$
252 trading days. Theory shows that the distribution of
$R_{T,0}=\Sigma_{t=0}^{T-1}s_{t}$ is then normal with drift
parameter $\Sigma\mu_{t}=T\bar{\mu}$ and width parameter
$\sqrt{\Sigma \sigma_{t}^{2}}.$ The Black-Sholes price is based on
hedging. Then there is a natural `short' time which disappears in
the continuous version of the theory. This fundamental short  time
is the interval between rehedgings. If this time is too short,
transaction costs become excessive. If it is too long, the expansion
given below breaks down. Although it is in reality more complicated,
assume for simplicity that the time between rehedging is $\Delta
t=1$ day. The dealer sells an option and maintains a hedge. Let the
number of hedge shares owned between time $t$, $t+1,$ be $\phi_{t}.$
Assume the risk free rate is zero, as it is known how to restore it
to the formulas by a trick at the end. Invoking the usual
no-arbitrage condition, the change in value of the hedge of the
dealer from time $t$ to time $t+1,$ but just before rehedging, is
equal to the change in his obligation to the buyer, i.e. to the
change of the fair value or price $C(S_{t,}t)$ of the option. Thus,
using a Taylor expansion to
calculate the change of price we have%
\[
\Delta S_{t}\phi_{t} =C(S_{t+1},t+1)-C(S_{t},t)\label{BS}\\
 \approx\Delta S_{t}\frac{\partial C}{\partial S_{t}}+\frac{1}{2}%
\frac{\Delta
S_{t}^{2}}{S_{t}^{2}}S_{t}^{2}\frac{\partial^{2}C}{\partial
S_{t}^{2}}+\frac{\partial C}{\partial t}+...
\]
Black-Sholes make the `delta' hedge $\phi_{t}=\partial C/\partial S_{t},$ and
choose the option price $C(S,t)$ as solution of the equation $\partial
C/\partial t=-\frac{1}{2T}\sigma_{BS}^{2}S^{2}\partial^{2}C/\partial S^{2},$
where obviously the `constant' $\sigma_{BS}^{2}/T$ should be given by
\begin{equation}
\sigma_{BS}^{2}=\Sigma_{t}\Delta S_{t}^{2}/S_{t}^{2}=\Sigma(\mu_{t}+\sigma
_{t}W_{t})^{2}\approx\Sigma\sigma_{t}^{2}=T\overline{\sigma_{t}^{2}%
}\label{sigBS}%
\end{equation}
if $T$ is large, and $\mu_{t}$ is small in the sense above. Note that for
large $T,$ $\sigma_{BS}^{2}\propto T.$ Also, the `miracle' has happened, the
option price does not depend on the drift parameter.

Furthermore, the future Black-Sholes volatility, (for $T\approx one$
$~year), $ is rather well predicted by the historical data. The
Black-Sholes volatility (divided by the horizon time) is and should
be, an average `daily' volatility, daily because of the `daily'
rehedging, and not necessarily the width of the distribution at time
$T,$ predicted by Eq. (\ref{bm}). Therefore, one has rather good
statistics, since there are many days until time $T+t$ is reached.
However, if $\mu_{t}$ is some `known' simple function of $t,$ which
maintains the smallness constraints, Eq.(\ref{bm}) does predict a
width $\sigma_{BS}.$

Let us next suppose that $\mu_{t}$ has a mean reversion property. Namely,
let us suppose that $\mu_{t}$ `points' to some future price goal
$S_{G}(t+1/\gamma)$, at a time $1/\gamma$ in the future. It points in the
sense that $S_{t}e^{\mu_{t}/\gamma}=S_{G}(t+1/\gamma)$ or $\mu_{t}=\gamma(\ln
S_{G}-\ln S_{t}),$ Suppose that $S_{G}(t)=S_{G}e^{\mu t}.$ Then, Eq.(\ref{bm})
becomes%
\begin{equation}
s_{t}=-\gamma(\ln S_{t}-\ln S_{G}-\mu t)\Delta t+\mu+\sigma_{t}\sqrt{\Delta t}W_{t}\label{ou}%
\end{equation}
which is a simple trending Ornstein-Uhlenbeck process. It has three drift
related parameters, $\mu,$ $S_{G},$ and $\gamma.$ Both $S_{G}$ and $\mu$ are
difficult to extract from the data on a single stock. Let
\begin{equation}
\,q_{t}=\ln S_{t}/S_{G}-\mu t.\label{qt}%
\end{equation}
Note that one expects $\overline{\left\vert q_{t}\right\vert }\approx
\sigma_{t}/\sqrt{\gamma}$ so that for this process $\mu_{t}^{2}\approx
\gamma\sigma_{t}^{2}<<\sigma_{t}^{2}$ if $\gamma\approx1/250$, i.e. $\gamma$
is an inverse year. Thus, the Black-Sholes option price is not changed if the underlying process is mean reverting rather than a random walk,
and is again determined by the average daily variance \cite{LoWang,Grundy}.

The volatility of the future distribution predicted by the OU
process has a different value, however. It is given by
$\sigma_{OU}^{2}=\sigma_{BS} ^{2}\left(  1-e^{-\gamma T}\right)
/\gamma T.$ There is also a negative correlation between successive
returns, $E_{OU}(\left(  R_{T,t}-\mu T\right) \left(  R_{T,t-T}-\mu
T\right)  )=-\frac{1}{2}\sigma_{OU}^{2}\left( 1-e^{-\gamma T}\right)
$ \cite{LoWang}. [These results are easily obtained by use of an
explicit solution for
$q_{t}=\Sigma_{s=-\infty}^{t}\sigma_{s}e^{-\gamma\left( t-s\right)
}W_{s},$ which can be approximated as a stochastic integral. Because
we don't know $S_{G},$ we don't know the value of $q_{0}$ at time
$t=0$ even if we know the price $S_{0}$ at that time. Setting the
lower limit in the sum far into the past makes $q_{0}$ a stochastic
quantity and supplies a sort of average over $S_{G}$. The results
are different if $S_{G}$ and $S_{0}$ are known.]

Another model with some plausibility is one in which the drift is constant for
a while, probably of order a year, then in a short time of a month or two
jumps to a new constant. This process would have to be mean-reverting also.
Since it requires even more parameters and is not well worked out, we do not
discuss it further.

\section{Averaging over predicted mean returns}\label{sec:data}

Our idea is to assume an underlying random walk or trending OU process, where
the chief unknown parameter is the expected return $\mu_{t}$. If we knew this
parameter, the PDF would be log normal with the width dependent on the
process. Our recommendation to the subset of investors who try to estimate a
stock's future PDF, is that they average somehow over reasonable predictions
of the expected return, which will result in a distribution that is wider than
the underlying widths $\sigma_{BS},$or $\sigma_{OU.}$

There are many schemes utilized
to predict the future course of stocks. There are at least three general
methodologies. Value investors believe that such underlying factors as
earnings, book value, cash flow, can be predicted and can be theoretically
converted into a target price for the stock. Most research and commercial
forecasting schemes like this are based on factor betting in the spirit of
Fama and French \cite{FF1992,FF2000,Cooper,BARRA}. These schemes
tend to predict target prices, which are related to the underlying `intrinsic
value' of the stock.

Another methodology is that of technical investing. Its proponents believe
that various shapes of stock price history graphs give buy or sell signals. It
is difficult to turn this into a prediction of the drift parameter.

One well documented version of technical investing is trend following or momentum investing \cite{Cooper,Jagadeesh}. In that case it is assumed that what
happened to the stock in the recent past is likely to continue for a while.

A third technique tries to take advantage of mean reversion. For example,
one looks over a cross section of stocks and invests in those for which the
recent returns have been unusually low. A value version is that one chooses
stocks whose book to price ratio, say, is unusually high, in the expectation
that the price will rise to bring the ratio down to more normal
values \cite{DbT}.

It is apparent that not many of these techniques actually attempt to predict
the expected return. In any case, it is not easy to extract such predictions
from the published reports of analysts. The exception is momentum investing
which assumes the actual return of the recent past can be used to predict the future \cite{Cooper,Jagadeesh}.

\begin{figure}
\centerline{\epsfig{file=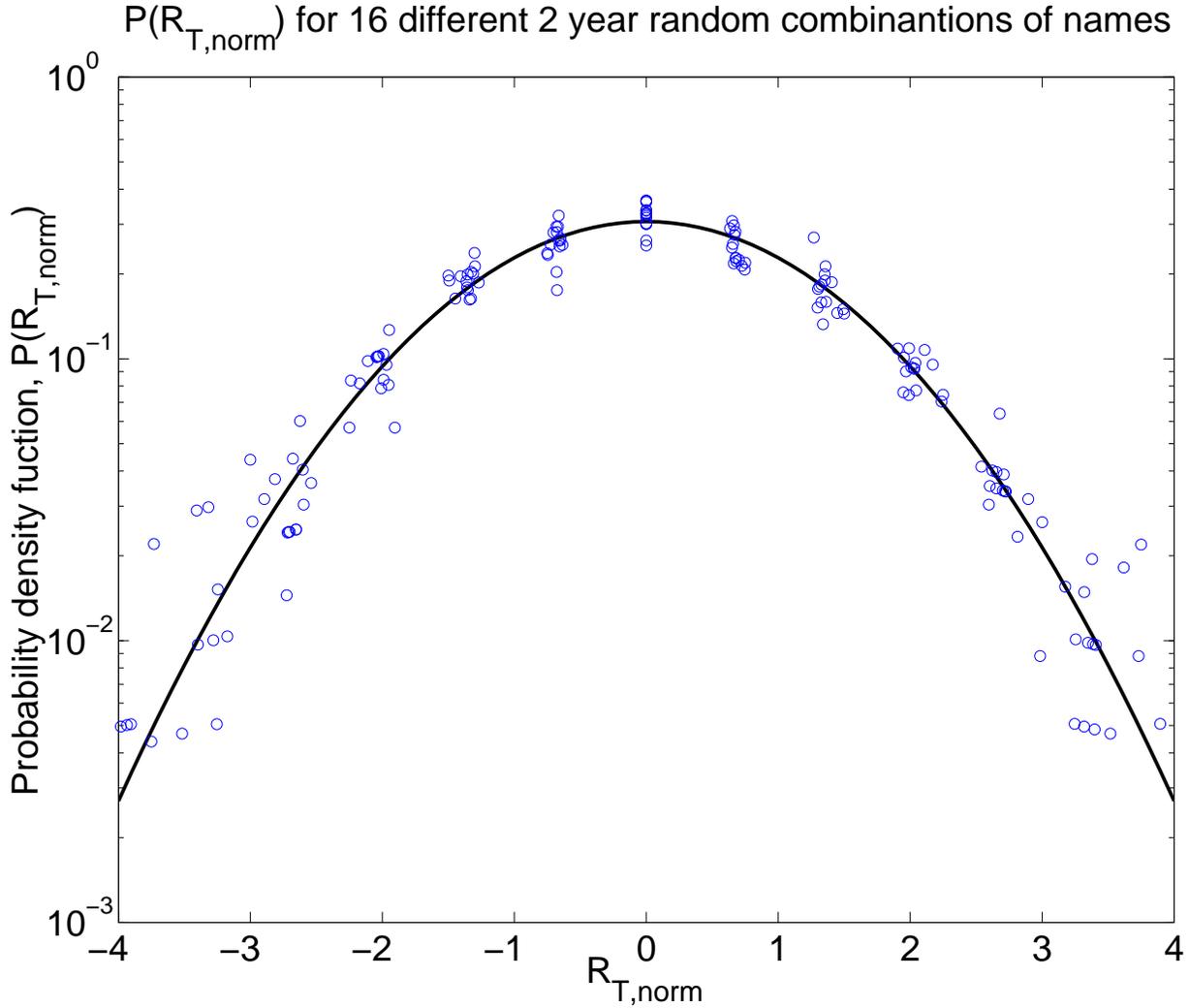,width=1\linewidth,angle=-90}}
\caption{\footnotesize\sf Empirical PDF for $304$ names from the
  S\&P500 where the correlation between names is removed. Two random
  years are chosen for each name. Different random choices result in
  slightly different PDFs. The solid black line represents the best fit gaussian. The mean is
  approximately zero and the standard deviation is $1.30\pm0.04$. We estimate the virtual volatility to be of the order of
  $26\%$ to $34\%$.}
\label{fig:mu.norm}
\end{figure}

We still face the fundamental difficulty, stressed above, that for a given
name, it is not possible to have enough exemplars of year-long price histories
to determine the parameters. We must therefore go to a cross-sectional
approach. Assume that we can choose a large number of similar names, which
for the present purpose differ only by their mean drift and volatility\cite{Stanley}. If
they are mean reverting, we assume their mean reversion parameters $\gamma$ are similar.

To compare one name with another, we normalize the returns. A natural first
guess is the $t$- or Student statistic,

\begin{equation}
R_{T,Stud}=\frac{R_{T,t}-\bar{\mu}T}{\sigma_{BS,T,t}} \label{En}%
\end{equation}
where $R_{T,t}=ln(S_{T+t}/S_{t})$ is the $T$ horizon return, the volatility
is that measured over the period $t,t+T,$ and $\bar{\mu}$ is the `known'
average drift parameter. If the returns are normal, the distribution of the
statistic of Eq.(\ref{En}) will be a Student distribution with $T$ degrees of
freedom. For large $T,$ this is nearly a normal distribution with zero mean
and unit width.

However, we want to make a prediction based on history to time $t,$ so we
replace the volatility in the denominator by the historical volatility.
[Other approximations for $\sigma_{BS}$ work well also.]

Most importantly, we replace $\bar{\mu}T$ by an estimate which has a
distribution. Namely we replace $\bar{\mu}T$ by the guess of the momentum
investor, $\bar{\mu}T\rightarrow R_{T,t-T}=ln(S_{t}/S_{t-T})$. The result is a
statistic which in effect \emph{averages} the predictions of the drift as a
momentum investor would do,
\begin{equation}
R_{T,norm}^{n}=\frac{R_{T,t}^{n}-R_{T,t-T}^{n}}{\sigma_{BS}^{n},_{t-T}}.
\label{Rname}%
\end{equation}
We add a label $n$ giving the name of the stock. Our interpretation
of this statistic is that its width estimates the enhancement factor
above the empirically known $\sigma_{BS}^{n}$ for the group of
stocks due to the virtual volatility effect.

To illustrate this empirically, we take sixteen years of daily data
ending Jan. 2006 of a $304$ stock subset of the S\&P500. The subset
was of stocks present in the S\&P over the chosen time-span. [We
have tried other time periods and other subsets with very similar
results,including data that spans from Oct. 1993 to Oct. 2002.] The
data was downloaded from Yahoo \cite{Yahoo}. We deliberately wanted
to have rather similar stocks, in this case rather large and
actively traded companies, so that a cross sectional analysis has
meaning.

One possible problem with the set of stocks we chose is that there
is known to be considerable temporal correlation between stock price
changes. Since studying the effects of correlation between names is
not our interest here, we chose, for each name, three two year
periods, at random, from the data. In other words, we took $T=252$
in Eq.(\ref{Rname}) and $t$ was chosen at random in the interval
Jan. 1990-Jan. 2006. This gives a sample of size $2430$ from which
to estimate the distribution.

The distribution is \emph{normal} with
essentially \emph{zero mean} and a \emph{width }(standard deviation) about $1.26-1.34$ (Fig. \ref{fig:mu.norm}). The
normality of this distribution is consistent with the argument that
year-in-the-future stock price distributions are lognormal. The zero mean is
evidence that yearly past returns are unbiased predictors of yearly future
returns. The width, 1.26-1.34, says that the volatility $\sigma_{T}$ of a given
stock, a year in the future, is, on average, 26\% - 34\% greater than the empirical
historical volatility for that stock. The reason for this
increase in width or decrease in predictability is that we don't know the
`true' drift parameter, but have effectively averaged it as a momentum
investor would.

\section{Mean Reversion}

This result is the main result of this paper. However, the statistic defined
by Eq.(\ref{Rname}) has some nice features, even without the virtual
volatility interpretation. This is especially true when $T$ is large enough
that fluctuations in the empirically obtained denominator can be ignored or
taken into account as a perturbation. One virtue, in contrast to the standard
regression techniques used to study mean reversion, is that the double
difference structure of the numerator of the statistic Eq.(\ref{En}) is
independent of drift parameters, known or unknown, if they are constant.

Let us assume first, that the $2430$ samples are generated by a random walk of
Eq.(\ref{bm}) with constant drift parameters $\mu^{n}$, i. e. constant over
two years for each choice of name and time interval. In each exemplar we may
add and subtract the correct (even though unknown) $\mu^{n,t}$ in the
numerator to find that such a random walk statistic theoretically has mean
zero and variance \emph{two.} In other words, a standard ensemble of random
walks, but with \emph{unknown} drift parameters, leads to a virtual volatility
enhancement of $\sqrt{2}-1\approx42\%.$ [Because of small fluctuations in the
denominator, it is actually slightly larger.]

Real stocks are therefore \emph{more predictable }than the random walk with
constant but unknown parameters itself. The reason is that real stocks are
mean reverting \cite{DbT,FF1988,LoMac}. Some argue that mean reversion contradicts the idea of market efficiency, therefore it is sometime termed an anomaly \cite{DbT}.

Suppose we assume that the stocks of Eq.(\ref{Rname}) are mean reverting with
the simple Ornstein-Uhlenbeck trending process. We take the reversion rate
$\gamma$ constant over all time and all names. We also assume constant but unknown
drift $\mu^{n}$ and fixed price target $S_{G}^{n}$ for each two year interval
and name. The numerator of Eq.(\ref{Rname}) can be expressed in terms of the
normalized parameter $q_{t},$ Eq.(\ref{qt}), without needing to know the drift
and target price. The result is:
\begin{equation}
var(R_{T,norm})=\left[  \frac{(1-e^{-\gamma T})}{\gamma T}\right]  \left[
2+\left(  1-e^{-\gamma T}\right)  \right]  \label{vOU}%
\end{equation}

\noindent The first square bracket is the ratio of the Ornstein-Uhlenbeck
variance of $R_{t\pm T}^{n}$ to the Black-Sholes volatility-squared of the
same quantity. The parenthesis in the second square bracket comes from the
negative correlation between history year and predicted year, as is
characteristic of mean reversion.

From the numerical result for actual S\&P stocks, we can conclude
that an average $\gamma^{-1}$ is about 1.1 - 1.3 years for $T=1$
year, that is, the target price is a little above 1 year in the
future. Assuming the same OU process applied to Fama and French
results\cite{FF1988}, we arrive at $1.9$ to $3.1$ years for small
cap stocks and $4.3$ to $7.2$ years for large cap. Conversely our
results indicate a correlation of one year returns of -0.32 (Fama
and French find an insignificant correlation for one year returns).
Our results are not in agreement with Fama and French, however we do
not try to compare both. We have used different samples, different
time periods and moreover we have assumed a very simple mean
reversion model. We are only drawing the attention to the order of
magnitude agreement we have achieved with a very crude model. If one
takes seriously the OU model and formula \ref{vOU}, then the virtual
volatility effect is enhanced, namely the actual width
$\sigma_{T}\approx1.5\sigma_{OU},$ for $T=1$ year.

However, real stocks are not so simple. Consider the empirical distribution
of Eq.(\ref{Rname}) for different values of $T.$ The results are given in
Table \ref{table:gammas}.

\begin{table}
\caption{\footnotesize\sf 1. The empirical estimate of the ratio of standard
deviation of normalized log returns to the Black Sholes volatility, as a
function of time horizon. 2. The rate of reversion to the mean calculated from
the empirical estimate on the assumption that the stocks follow a simple
trending OU process. \label{table:gammas}}
\begin{tabular}
[c]{llllllllll}%
T(quarters) & 1/3 & 1 & 2 & 3 & 4 & 5 & 6 & 7 & 8\\
$\sigma_{T}/\sigma_{BS}$ & 1.49 & 1.37 & 1.32 & 1.26 & 1.33 & 1.35 & 1.37 &
1.46 & 1.49\\
$\gamma^{-1}(years)$ & ** & 7.7 & 2.4 & 1.15 & 1.3 & 1.2 & 1.3 & ** & **
\end{tabular}
\end{table}

The notation ** means that the simple OU process cannot give the result.
Indeed, we can conclude that although the stocks are mean reverting, there
is considerably more structure than found in the simplest OU process.

Since mean reversion is not the main subject of this paper, we confine
ourselves to a simple, qualitative explanation of the table as follows.
There is more than one time scale of stock price temporal correlation. In
addition to the negative correlation at longer time \cite{DbT,FF1988}, there is a
shorter time positive correlation at horizons of order week-month
\cite{Lo,DbT,LoMac}.

At still longer horizons, we suggest that the approximation that the drift
$\mu_{t}$ is constant breaks down. If the drift is different for $R_{t+T}$ and
$R_{T-t},$ there is a positive addition to the variance. Thus, $\dot{\sigma
}_{T}/\sigma_{BS}$ has a minimum. In fact, for horizons of 7 and 8 quarters,
the mean of $R_{T,norm}$ becomes distinctly negative. That means that the
mean drift at the earlier time is on average greater than the mean drift at
the later time. In other words, the returns earlier in the period were greater
than the returns later in the period. Certainly for the market as a whole,
during this period prices first went up during the '90's and turned over at
about year 2000.

\section{Investment consequences of the virtual volatility}

We have just recounted arguments that the expected return parameter for a
given stock or portfolio is not known very well and its value is not agreed on
by market participants. We do not discuss the consequences of this for such
concepts as the `efficient frontier' of modern portfolio theory [MPT]
\cite{Bodie,Markowitz,Markowitz1979,Huang}.

We argued that one way of thinking about this lack of knowledge and agreement
on the expected return is that the PDF of future price distributions is wider
than usually assumed, something like 1.3$\sigma_{BS}$. What are some of the
consequences for investors? Since volatility is equated to risk in typical
textbook finance, we see that stocks are risker by 30\% than we thought they were. A
favorite single number that tries to summarize how well a portfolio is going
to perform, in the risk-return sense,  is the \emph{ex ante }Sharpe ratio,
$Sr=\left(  \mu_{T}-rT\right)  /\sigma_{T}$. This number tries to balance the
risk, defined as the volatility, with the return in excess of the risk free
return. Obviously, it is reduced by 30\% or so as compared with the ratio
using the volatility $\sigma_{BS}.$ Going a bit beyond this is MPT, which
uses of the utility functional $U(I_{0})=I_{0}(\mu_{T}-rT)-\frac{1}{2}%
AI_{0}^{2}\sigma_{T}^{2}.$ Here $A$ is the investor's risk aversion parameter.
Minimizing $U$ with respect to the investment amount, $I_{0},$ gives
$I_{0}=\left(  \mu_{T}-rT\right)  /A\sigma_{T}^{2}$ which means that an
investor using this MPT utility, would invest a factor 1.7 less because of
the virtual volatility effect. The utility itself is changed to $U=\frac
{1}{2A}Sr^{2}$, so the utility of the investment is reduced by the same rather
significant factor.

It must be admitted that very few investors, (probably including Sharpe) know
their personal value of the risk aversion parameter $A,$ so, all that needs to
be done is to have everyone decrease their $A$ by a factor 1.7, in other
words, become less risk averse than they had thought, and nothing would
change. Thus, even though the previous calculation (without the virtual
volatility effect) is in several textbooks, it is doubtful that it has much
effect on live investors.

Nevertheless, it is clear and in agreement with intuition and standard lore,
that \emph{additional uncertainty makes an investment less attractive.}

There are, however, other ways of investing in stocks or indices than just
buying them. One of the simplest is investing in calls on the underlying
stock. In this paper we confine ourselves to pointing out the following. The
expected excess gain, as predicted at the time of making the investment, for
the strategy of buying a call at price $C_{K}$ and holding it until expiration
at a time $T+t$ in the future, is given by
\begin{equation}
E_{t}[G]=I_{0}e^{rT}\left(  \frac{E_{t}\left(  \left(  S_{T+t}-K\right)
^{+}\right)  }{C_{K}e^{rT}}-1\right)  \label{EGainCall}%
\end{equation}
Here $I_{0}$ is the amount invested, $S_{T+t}$ is the price of the underlying
at expiration, and $K$ is the strike price. The expectation
\begin{equation}
E\left[  \left(  S_{Tt}-K\right)  ^{+}\right]  =\int_{K}^{\infty}%
dS(S-K)p_{t}(S,\tilde{\mu}_{T},\sigma_{T})\label{excall}%
\end{equation}
where $p_{t}(S_{T+t},\tilde{\mu}_{T},\sigma_{T})$ is the \emph{ex-ante,
}predicted, log-normal distribution of final stock price that we have just
discussed. The Black-Sholes call price \cite{BS,Bjork}, (approximately
$C_{K}),$ is given by the same formula, Eq. (\ref{excall}) with $p$ replaced
by the risk neutral world distribution, $p(S_{T+t},rT,\sigma_{BS}).$ The
actual call price is subject to various small corrections that are summarized
by the replacement of $\sigma_{BS}$ by the implied volatility $\sigma_{I}.$
Thus $E(G)$ vanishes if $\tilde{\mu}_{T}=rT$ and $\sigma_{T}=\sigma_{I}.$[Note
that Eq.(\ref{EGainCall}) for $K=0$ gives the gain from buying the underlying
when $C_{K=0}$ is the price of the stock, $S_{t}$.]

The expected gain is an increasing function of three parameters. The first,
obviously, is the drift parameter $\tilde{\mu}_{T}.$ The second is the strike
price $K.$ Both the numerator and denominator in Eq.(\ref{EGainCall}) decrease
with increasing $K,$ but the denominator decreases faster. [For options very
much out of the money, $C_{K}$ is so small as predicted by Black-Sholes that
the buy-sell option price spread starts to dominate.] The result is that
somewhat out-of-the-money calls on stocks with projected good returns have
considerably higher expected return than does the underlying itself, even if
the volatility $\sigma_{T}$ is mistakenly kept at $\sigma_{I}$.

The third parameter is the virtual volatility $\sigma_{T}.$ As $\sigma_{T}$
increases, the expected return increases. The expected final call value of
Eq.(\ref{excall}) increases, the more so for out of the money calls. The cost
$C_{K}$ is independent of predictions of the future return. The wider virtual
tails of $p(S_{T},\tilde{\mu}_{T},\sigma_{T})$ on the up side increase the
return, while the losses coming from low side tails are limited to the cost of
the call. The virtual volatility effect thus increases the expected return on
call buying, without much increasing the prospect of losses. Similarly, if the
actual distribution has fatter tails than the log-normal that we have assumed,
it also raises the expected return without increasing much the prospect of
losses. In fact, for out-of-the-money calls, there is a significant expected
gain even if the return $\tilde{\mu}_{T}$ vanishes.

There is some empirical evidence which tends to support these
ideas \cite{Shumway,Rendleman}. That work of course has nothing to do with
prediction. It did however find that the historical mean returns on certain
rather short term ($\approx$ one month) calls was impressively high, the more
so for out of the money calls.

A complete analysis requires a discussion of risk or utility in call-buying, a
situation for which the `risk = volatility=standard deviation' formulation of
traditional MPT is far from adequate \cite{Madan2001,Ingersoll1982}. We defer
this discussion to a future publication. However, for those investors who are
mainly averse to losing too much money, [as opposed to being averse to both
upside and downside uncertainty equally,] it is quite clear that the virtual
volatility effect as applied to calls\emph{\ increases} the desirability of
call buying, contrary to the outcome when purchase of the underlying is contemplated.

In other words, \emph{this is a case where certain rational investors can and
should regard uncertainty and inability to predict correctly as \textbf{GOOD}!
}

We thank V. M. Yakovenko for several suggestions. We thank J-P.
Bouchaud for pointing out Ref. \cite{DbT}.

\end{document}